\documentclass[a4paper,fleqn]{cas-dc}

\usepackage[numbers]{natbib}
\usepackage{amsmath,amssymb,amstext}
\usepackage{graphicx,hyperref}
\usepackage{multirow} 
\usepackage{amssymb}
\usepackage{enumitem}
\usepackage{bbm}
\usepackage{booktabs}
\usepackage{subfigure}
\usepackage{balance}
\usepackage{longtable}
\def\tsc#1{\csdef{#1}{\textsc{\lowercase{#1}}\xspace}}
\tsc{WGM}
\tsc{QE}
\tsc{EP}
\tsc{PMS}
\tsc{BEC}
\tsc{DE}

\begin{document}
\let\WriteBookmarks\relax
\def\floatpagepagefraction{1}
\def\textpagefraction{.001}
\shorttitle{Graph-Guided Structured Recovery of Modulo EEG Signals}
\shortauthors{Hazra et~al.}

\title [mode = title]{Graph-Guided Structured Recovery of Modulo EEG Signals}                      



\author[1]{Soujanya Hazra}[
                        orcid=0009-0001-4477-6818]

\credit{Conceptualization, Methodology, Software, Validation, Visualization, Writing - original draft}


\author[1]{Sanjay Ghosh}[
                        orcid=0000-0002-0474-5072]
\cormark[1]
\ead{sanjay.ghosh@ee.iitkgp.ac.in}

\credit{Conceptualization, Investigation, Funding acquisition, Supervision, Resources, Project administration, Writing - review and editing}

\affiliation[1]{organization={Department of Electrical Engineering, Indian Institute of Technology Kharagpur},
                postcode={721302}, 
                state={WB},
                country={India}}



\cortext[cor1]{Corresponding author}

\begin{abstract}
Electroencephalography (EEG) acquisition is often affected by large-amplitude variations across subjects and sessions. This makes conventional front ends vulnerable to saturation and clipping. Modulo sampling addresses this limitation by folding the signal into a bounded range rather than truncating it. The recovery task is then to reconstruct the original waveform from periodic observations, which is a highly ill-posed inverse problem. In this work, we propose a compact signal-structured framework for modulo EEG recovery. 
The method combines phase-aware features, graph-temporal modeling, and structured fold-state decoding. A boundary-guided gating technique improves inference near wrap events, where uncertainty is highest.
Experiments across two EEG datasets, over a wide range of folding thresholds, demonstrate consistent recovery performance. Ablation and noise studies further support the contribution of the structured decoding technique. 
The proposed signal-processing approach effectively recovers folded EEG recordings. These results show that structured fold-state decoding is effective for recovering folded EEG.
The code is accessible at \url{https://github.com/soujo/graphunwrapnet}.
\end{abstract}

\maketitle

\section{Introduction}
\label{sec:intro}
Electroencephalography (EEG) is widely used in brain-computer interfaces because it is non-invasive, portable, and has high temporal resolution \cite{wolpaw2000bci,Abiri2019bci}. A practical difficulty, however, is the large amplitude variation across subjects, sessions, and tasks. This variation makes EEG acquisition vulnerable to saturation when the analog front end operates within a limited dynamic range. Once clipping occurs, the lost information cannot be recovered.
Modulo sampling provides an alternative to clipping by folding the signal into a bounded interval \cite{Bhandari2017usf, Bhandari2021usr, romanov2019nyquist}. For a latent EEG sample $x$ and folding threshold $\lambda$, the modulo observation $p$ satisfies:
\begin{equation}
x = \lambda z + p, \qquad p \in [0,\lambda),
\label{eq:modulo}
\end{equation}
where $z \in \mathbb{Z}$ is the unknown fold state. The recovery task is to reconstruct $x$ from $p$ without direct access to $z$. This is a highly ill-posed inverse problem, since multiple unfolded candidates can map to the same wrapped observation. Similar ambiguity has also been studied in modulo compressed sensing and sparse recovery settings \cite{prasanna2021identifiability, shah2021sparse}.

Modulo recovery has been studied in several signal processing settings. Unlimited sensing established the basic sampling and reconstruction theory for folded measurements \cite{Bhandari2017usf,Bhandari2021usr}. 
Recent work also studied robust, hardware-oriented modulo recovery \cite{yan2025difference}.
Related ideas have also appeared in modulo imaging, where wrapped intensities are reconstructed for high-dynamic-range acquisition \cite{zhao2015MRF,Shah2017recon}. For graph signals, folded recovery has been formulated using integer and sparse optimization tools \cite{feng2022SparseOptimization}. In EEG, recent learning-based work has shown that deep models can improve modulo recovery \cite{geng2023transformer}. 
Despite these developments, existing methods do not explicitly combine the temporal evolution of the hidden fold state with electrode-topology-aware representation learning. Optimization-based graph recovery methods exploit inter-node relationships, but generally do not learn signal-dependent temporal representations. Conversely, learning-based modulo EEG recovery methods model temporal dependencies, but do not explicitly enforce global consistency of the discrete fold-state trajectory. These limitations become increasingly important as the folding threshold decreases and the number of admissible unfolded candidates increases.
These observations motivate a more structured formulation. Modulo EEG recovery is not only a regression problem. It is also a latent state estimation problem. The hidden fold trajectory must be inferred from periodic measurements, while preserving temporal consistency and inter-channel dependence. This suggests three natural priors: \emph{i)} wrapped-domain features should reflect modulo periodicity, \emph{ii)} channel interactions should follow the EEG graph, and \emph{iii)} fold transitions should obey a structured temporal model.

Accordingly, the proposed framework treats modulo EEG recovery as a structured fold-state estimation problem that combines periodic feature representation, graph-temporal modeling, and conditional random-field decoding. The main contributions are as follows:
\begin{enumerate}
    \item We reformulate the modulo EEG recovery as hidden-fold-state inference with conditional random-field decoding of local increments and global states.
    \item We combine phase-aware wrapped-domain features with graph-temporal representation learning to exploit modulo periodicity and inter-channel dependence.
    \item We incorporate boundary-guided transition sharpening, residual refinement, and lightweight calibration to improve stability near wrap events.
    \item We validate the proposed framework on the STEW \cite{lim2018STEW} and MIMED \cite{wirawan2024mimed} datasets over a broad range of folding thresholds, and further examine its behavior through ablation and noise-robustness studies.
\end{enumerate}

The paper is organized as follows. 
Section~\ref{sec:related_work} reviews current methodologies. Section~\ref{sec:method} presents the proposed structured recovery framework. Section~\ref{sec:exp} presents the experimental setup, reconstruction results, robustness analysis, and ablation studies.
Section~\ref{sec:conc} concludes our work.

\section{Related Work} \label{sec:related_work}

Modulo EEG reconstruction is related to three main research directions: modulo sampling, graph-based signal recovery, and learning-based EEG modeling. This section briefly reviews these directions and identifies the limitations addressed by the proposed framework.

\subsection{Modulo Sampling and Signal Recovery} 

Modulo sampling avoids saturation by folding amplitudes into a bounded interval and recovering the original signal through unknown integer fold values. Unlimited-sampling theory established recovery conditions for bandlimited signals using finite differences and sufficient oversampling~\cite{Bhandari2017usf, Bhandari2021usr}. Romanov and Ordentlich further showed that modulo folding does not fundamentally increase the sampling-rate requirement above the Nyquist rate \cite{romanov2019nyquist}. Later methods improved robustness through residual estimation \cite{azar2022residual} and considered sparse recovery from modulo measurements \cite{prasanna2021identifiability,shah2021sparse}. Similar periodic inverse problems have also been studied in high-dynamic-range imaging \cite{zhao2015MRF}. These approaches rely mainly on bandlimitedness, sparsity, or other predefined signal priors. Such assumptions do not fully represent the nonstationary temporal structure and inter-subject variability of EEG.


\subsection{Graph and Learning-Based EEG Recovery} 
EEG channels exhibit spatial and functional dependence, motivating graph-based processing. Ji et al.\ formulated modulo recovery for graph signals using graph-bandlimitedness and integer or sparse optimization \cite{feng2022SparseOptimization}. Their work showed that inter-node relationships can reduce folding ambiguity, but the method does not learn EEG-specific temporal representations. Graph neural networks have also been effective in EEG analysis by propagating information between related electrodes \cite{Song2020emotion}. However, most existing EEG graph models address classification rather than waveform reconstruction. For modulo EEG, Geng et al.\ proposed a transformer-based recovery model that learns long-range temporal dependencies from folded observations \cite{geng2023transformer}. Although data-driven modeling improves recovery, direct waveform prediction does not explicitly enforce consistency of the latent integer fold sequence or incorporate electrode topology. An isolated fold error may therefore shift the reconstructed signal by one or more folding intervals.

\subsection{Structured Recovery} 
Modulo EEG reconstruction can be viewed as discrete fold-state estimation followed by waveform recovery. Since fold states are typically piecewise constant and change near wrap boundaries, adjacent states are strongly dependent. Conditional random fields (CRF) provide a suitable mechanism for combining local state evidence with transition constraints and decoding a globally consistent sequence \cite{lafferty2001crf}. 
The existing literature therefore leaves three complementary limitations: \emph{(i)} Classical modulo-recovery algorithms exploit analytic signal priors but do not learn EEG-specific representations. 
\emph{(ii)} Folded graph-signal methods use inter-node structure but rely mainly on optimization and graph-bandlimitedness. 
\emph{(iii)} Learning-based modulo EEG methods capture temporal patterns but do not explicitly enforce electrode-topology-aware processing or global consistency of the discrete fold trajectory. 

To address these limitations, the proposed framework formulates modulo EEG reconstruction as structured fold-state inference. Periodic phase features represent the circular geometry of modulo observations, while a graph-temporal backbone jointly models temporal morphology and electrode dependence. The network estimates both absolute fold states and local fold increments, and a linear-chain CRF combines this evidence to decode a globally coherent fold trajectory. Boundary-guided transition sharpening emphasizes uncertain wrap locations, while bounded residual refinement corrects small reconstruction errors without replacing the discrete fold-state estimate.

\section{Methodology: Fold-State Recovery}
\label{sec:method}

We formulate the modulo EEG recovery as a structured estimation of a hidden fold-state sequence. This viewpoint differs from direct unfolded-signal regression and is better aligned with the modulo sensing model \cite{Bhandari2017usf,Bhandari2021usr}. It also allows local wrap transitions and global fold consistency to be handled jointly.
\subsection{Observation Model}

Consider a subject-wise normalized EEG segment $\mathbf{x}\in\mathbb{R}^{T\times C}$ with $T$ time samples and $C$ channels. To reduce inter-subject amplitude variation, each channel is first normalized using a median/MAD transform, followed by sigmoid compression:
\begin{equation}
\tilde{x}_{t,c}
=
\sigma\!\left(
\alpha \frac{x_{t,c}-m_c}{d_c+\epsilon}
\right),
\label{eq:robust_norm}
\end{equation}
where $m_c=\operatorname{med}_{t}(x_{t,c})$ is the channel-wise median, $d_c=\operatorname{med}_{t}(|x_{t,c}-m_c|)$ is the corresponding median absolute deviation (MAD), $\alpha>0$ is a scaling parameter, and $\epsilon>0$ is a small constant for numerical stability. For notational simplicity, we use $x_{t,c}$ to denote the normalized sample hereafter.
Under modulo acquisition, the observed folded sample $p_{t,c}$ satisfies:
\begin{equation}
p_{t,c}=x_{t,c}\bmod \lambda,
\qquad
x_{t,c}=\lambda z_{t,c}+p_{t,c},
\qquad
p_{t,c}\in[0,\lambda),
\label{eq:mod_model}
\end{equation}
where $\lambda$ is the folding threshold and $z_{t,c}\in\mathbb{Z}_{\ge 0}$ is the unknown fold count \cite{Bhandari2017usf,Bhandari2021usr}. Since the normalized signal lies in $(0,1)$, the fold count takes values in a finite set, namely
\begin{equation}
z_{t,c}\in\mathcal{Z}=\{0,1,\dots,S-1\},
\qquad
S=\left\lfloor \frac{1}{\lambda}\right\rfloor+1.
\label{eq:state_space}
\end{equation}
To characterize temporal wrap evolution, we further define the fold increment as:
\begin{equation}
k_{t,c}=z_{t,c}-z_{t-1,c}, \qquad t=2,\dots,T,
\label{eq:k_def}
\end{equation}
and set $k_{1,c}=0$ for initialization. The variable $k_{t,c}$ localizes wrap transitions, whereas $z_{t,c}$ determines the absolute unfolding level. In contrast to optimization-based folded graph recovery \cite{feng2022SparseOptimization} and recent transformer-based modulo EEG recovery \cite{geng2023transformer}, the proposed framework estimates both quantities jointly within a structured sequence model.
\subsection{Phase-Aware Wrapped-Domain Features}

The wrapped signal is periodic in nature, so its representation should reflect modulo geometry rather than raw amplitude alone. For each $(t,c)$, we form:
\begin{equation}
\mathbf{f}_{t,c}=
\big[p_{t,c},\,\Delta p_{t,c},\,b_{t,c},\,\sin\phi_{t,c},\,\cos\phi_{t,c}\big]^{\top},
\label{eq:feat}
\end{equation}
where $\Delta p_{t,c}=p_{t,c}-p_{t-1,c}$, $b_{t,c}=\min(p_{t,c},\lambda-p_{t,c})$, and $\phi_{t,c}=2\pi p_{t,c}/\lambda$. The sinusoidal pair embeds the periodic observation model, while $b_{t,c}$ measures proximity to a wrap boundary. This is precisely the region where modulo inversion is most ambiguous, as also seen in modulo imaging \cite{zhao2015MRF} and periodic reconstruction problems \cite{Shah2017recon}.
\subsection{Graph-Temporal Representation Learning}

EEG channels are not independent. Their interactions are shaped by sensor geometry and correlated neural activity, which motivates graph-based coupling \cite{Bronstein2017geoDL,kipf2017semisupervised,Song2020emotion,ghosh2023gcn}. 
We model the EEG electrodes ($C$)  as the vertices of a channel graph $\mathcal{G}=(\mathcal{V},\mathcal{E})$, where $\mathcal{V}=\{1,\dots,C\}$ and each vertex corresponds to one channel. The edge set $\mathcal{E}$ is constructed from the electrode positions using a $\kappa$-nearest-neighbor rule, so that spatially neighboring electrodes are connected.
This yields a binary symmetric adjacency matrix $\mathbf{A}\in\{0,1\}^{C\times C}$. Self-loops are added through $\mathbf{A}+\mathbf{I}$, and $\mathbf{D}$ denotes the corresponding diagonal degree matrix. The normalized graph operator is then $\bar{\mathbf{A}}=\mathbf{D}^{-1}(\mathbf{A}+\mathbf{I})$.
The input feature is first mapped to a hidden representation through a learned linear projection,
\begin{equation}
\mathbf{h}^{(0)}_{t,c}=\mathbf{W}_{\mathrm{in}}\mathbf{f}_{t,c},
\end{equation}
where $\mathbf{W}_{\mathrm{in}}\in\mathbb{R}^{H\times 5}$ and $H$ the hidden dimension.
Each layer then combines dilated temporal convolution with graph mixing:
\begin{equation}
\mathbf{u}^{(\ell)}=\mathrm{Conv}_{d_{\ell}}\!\left(\mathbf{h}^{(\ell)}\right),\qquad
\mathbf{h}^{(\ell+1)}=\mathbf{h}^{(\ell)}+\varphi\!\left(\bar{\mathbf{A}}\mathbf{u}^{(\ell)}\right),
\label{eq:backbone}
\end{equation}
where $\mathrm{Conv}_{d_{\ell}}(\cdot)$ is a temporal convolution with dilation $d_{\ell}$, and $\varphi(\cdot)$ denotes  nonlinearity.
We refer to the graph-coupling step $\bar{\mathbf{A}}\mathbf{u}^{(\ell)}$ as a GraphMix operation, since it propagates features across neighboring electrodes at each time index.
The multiplication by $\bar{\mathbf{A}}$ is performed across channels at each time index. Dilated temporal convolutions provide an efficient sequence model with long effective memory \cite{bai2018tcn}.
From the final representation $\mathbf{h}^{(L)}_{t,c}$, the network predicts four quantities:
\begin{align}
\boldsymbol{\ell}^{(k)}_{t,c} &= \psi_k\!\left(\mathbf{h}^{(L)}_{t,c}\right)\in\mathbb{R}^{2K+1}, \\
\boldsymbol{\ell}^{(z)}_{t,c} &= \psi_z\!\left(\mathbf{h}^{(L)}_{t,c}\right)\in\mathbb{R}^{S}, \\
g_{t,c} &= \sigma\!\left(\psi_g\!\left(\mathbf{h}^{(L)}_{t,c}\right)\right)\in(0,1), \\
r_{t,c} &= \rho\lambda\tanh\!\left(\psi_r\!\left(\mathbf{h}^{(L)}_{t,c}\right)\right), \label{eq:rtc}
\end{align}
where $\boldsymbol{\ell}^{(k)}_{t,c}$ are increment logits, $\boldsymbol{\ell}^{(z)}_{t,c}$ are fold-state logits, $g_{t,c}$ is a boundary gate, $\rho>0$ controls the residual scale, and $r_{t,c}$ is a bounded residual term.
In addition, a feature-wise linear modulation (FiLM) block, driven by the channel-wise mean and standard deviation of the wrapped segment, is used to improve robustness across recording conditions.
\subsection{Structured Fold-State Decoding}

For each channel, the fold-count sequence is decoded by a linear-chain conditional random field (CRF) \cite{lafferty2001crf}. Let $U_t(i)$ denote the unary score for state $i\in\mathcal{Z}$ obtained from $\boldsymbol{\ell}^{(z)}_{t,c}$. The score of a candidate path $\mathbf{z}_{1:T,c}$ is:
\begin{equation}
\mathcal{S}(\mathbf{z}_{1:T,c})=
\sum_{t=1}^{T} U_t(z_{t,c})
+
\sum_{t=2}^{T} T_t(z_{t-1,c},z_{t,c}),
\label{eq:path_score}
\end{equation}
where the transition score is derived from the increment logits $\boldsymbol{\ell}^{(k)}_{t,c}$. Specifically, for a jump $j-i$ satisfying $|j-i|\le K$,
\begin{equation}
T_t(i,j)=\ell^{(k)}_{t,c}(j-i+K)-\beta\,\mathbbm{1}[i\neq j],
\label{eq:transition}
\end{equation}
and $T_t(i,j)=-\infty$ otherwise. 
Here $\mathbbm{1}[\cdot]$ is the indicator function, and $\beta>0$ is a Potts prior that discourages unnecessary state changes. This is appropriate because the fold trajectory is piecewise constant in time, with sparse wrap transitions.
Near wrap boundaries, local ambiguity is strongest. We therefore introduce a pre-estimation-guided feature injection (PGFI) gate $g_{t,c}$ that provides a boundary-aware confidence cue for transition sharpening. We use this gate to sharpen transition evidence by temperature scaling,
\begin{equation}
\boldsymbol{\ell}^{(k)}_{t,c}\leftarrow \boldsymbol{\ell}^{(k)}_{t,c}/\tau_{t,c},
\qquad
\tau_{t,c}=\max\!\big(\tau_{\min},\,1-\eta g_{t,c}\big),
\label{eq:temp}
\end{equation}
where $\tau_{\min}>0$ prevents over-sharpening and $\eta>0$ controls the gate strength. Thus, transition decisions become more decisive when the model detects a likely wrap event. The final fold-state sequence is obtained by Viterbi decoding:
\begin{equation}
\hat{\mathbf{z}}_{1:T,c}
=
\arg\max_{\mathbf{z}_{1:T,c}\in\mathcal{Z}^{T}}
\mathcal{S}(\mathbf{z}_{1:T,c}).
\label{eq:viterbi}
\end{equation}
This decoding step couples local fold-increment evidence with global fold-state consistency over the entire sequence. As a result, isolated noisy transition predictions are suppressed, while plausible wrap events are preserved.

\begin{figure*}[t]
\centering
    \includegraphics[width=0.99\textwidth]{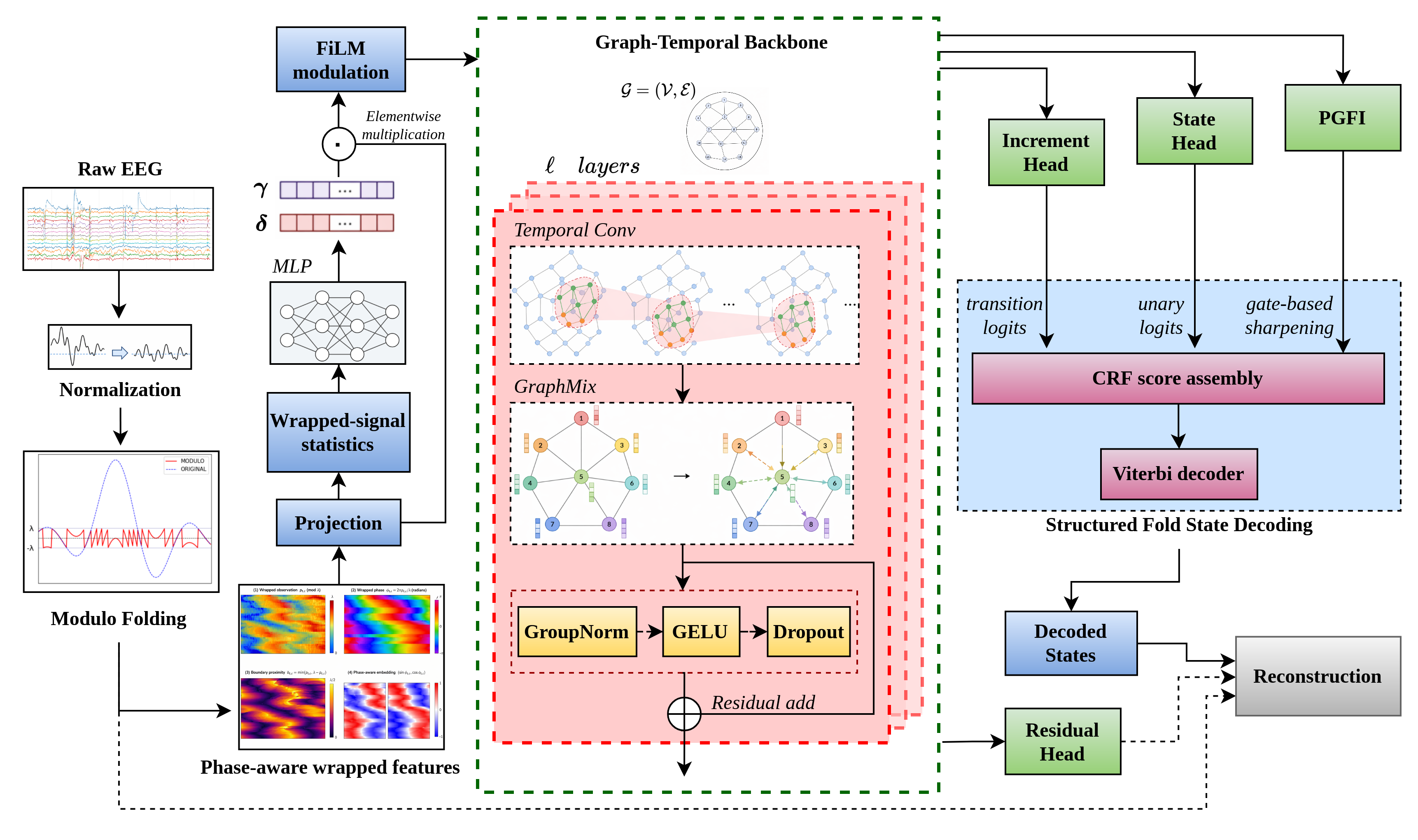}
\caption{Illustration of our proposed modulo EEG recovery framework.
Wrapped EEG is first converted to phase-aware features and processed by a graph-temporal backbone with GraphMix and FiLM calibration. The backbone predicts fold-increment logits, fold-state logits, a PGFI gate, and a residual term. A CRF with Potts prior decodes the hidden fold-state sequence, which is then used for waveform reconstruction.
}
\label{fig:graphunwrapnet}
\end{figure*}
\subsection{Reconstruction and Objective Function}

The unfolded EEG is reconstructed from the decoded fold-state sequence as:
\begin{equation}
\hat{x}_{t,c}^{\,\mathrm{hard}}=\lambda \hat{z}_{t,c}+p_{t,c}+r_{t,c}.
\label{eq:recon_hard}
\end{equation}
\noindent\textit{Fold-state consistency:}
From \eqref{eq:mod_model}, \eqref{eq:rtc}, and ~\eqref{eq:recon_hard},
\[
e_{t,c}\triangleq \hat{x}_{t,c}^{\,\mathrm{hard}}-x_{t,c}
=\lambda(\hat{z}_{t,c}-z_{t,c})+r_{t,c},
\qquad |r_{t,c}|\le \rho\lambda.
\]
Therefore, correct fold-state decoding implies
$|e_{t,c}|\le \rho\lambda$, whereas any incorrect fold state incurs
$|e_{t,c}|\ge (1-\rho)\lambda$. Hence, for $0<\rho<\tfrac12$,
correct and incorrect fold states are separated by a positive margin.
This consistency relation shows that reconstruction quality is directly controlled by fold-state accuracy.

\noindent\textit{Proof:}
Using
\[
x_{t,c}=\lambda z_{t,c}+p_{t,c}
\]
and
\[
\hat{x}_{t,c}^{\,\mathrm{hard}}=\lambda \hat{z}_{t,c}+p_{t,c}+r_{t,c},
\]
we obtain
\[
e_{t,c}=\lambda(\hat{z}_{t,c}-z_{t,c})+r_{t,c}.
\]
If $\hat{z}_{t,c}=z_{t,c}$, then
\[
e_{t,c}=r_{t,c},
\]
and therefore
\[
|e_{t,c}|\le \rho\lambda.
\]
If $\hat{z}_{t,c}\neq z_{t,c}$, then since $\hat{z}_{t,c}$ and $z_{t,c}$ are integers,
\[
|\hat{z}_{t,c}-z_{t,c}|\ge 1.
\]
Hence, by the reverse triangle inequality,
\[
|e_{t,c}|
\ge
\lambda |\hat{z}_{t,c}-z_{t,c}|-|r_{t,c}|
\ge
\lambda-\rho\lambda
=
(1-\rho)\lambda.
\]
Thus, for $0<\rho<\tfrac12$, correct and incorrect fold states are separated by a positive margin.

For differentiable training, we instead use the soft state estimate as:
\begin{equation}
\bar{z}_{t,c}=\sum_{s\in\mathcal{Z}} s\,\mathrm{softmax}\!\big(\boldsymbol{\ell}^{(z)}_{t,c}\big)_s,
\label{eq:z_soft}
\end{equation}
which gives the soft reconstruction:
\begin{equation}
\hat{x}_{t,c}^{\,\mathrm{soft}}=\lambda \bar{z}_{t,c}+p_{t,c}+r_{t,c}.
\label{eq:recon_soft}
\end{equation}
The overall training objective is:
\begin{equation}
\begin{split}
\mathcal{L} = {} & w_{\mathrm{crf}}\mathcal{L}_{\mathrm{crf}}
+ w_k\mathcal{L}_{k}
+ w_g\mathcal{L}_{g} \\
& + w_x\|\hat{\mathbf{x}}^{\,\mathrm{soft}}-\mathbf{x}\|_1
+ w_{\Delta x}\|\Delta\hat{\mathbf{x}}^{\,\mathrm{soft}}-\Delta\mathbf{x}\|_1
+ \mathcal{R},
\end{split}
\label{eq:loss}
\end{equation}
where $\mathcal{L}_{\mathrm{crf}}$ is the negative log-likelihood of the ground-truth fold path under ~\eqref{eq:path_score}, $\mathcal{L}_{k}$ is a boundary-weighted cross-entropy loss on the fold increment $k_{t,c}$, and $\mathcal{L}_{g}=\|g-q\|_2^2$ supervises the boundary gate using:
\begin{equation}
q_{t,c}=\max\!\left\{0,\min\!\left(1,\,1-\frac{2b_{t,c}}{\lambda}\right)\right\}.
\label{eq:q_target}
\end{equation}
Here $\Delta$ denotes the first-order temporal difference along time. The regularizer $\mathcal{R}$ contains small penalties on the residual and calibration parameters. The reconstruction terms promote waveform fidelity and temporal morphology, while the CRF term enforces consistency of the hidden fold-state trajectory. The resulting objective is therefore a structured signal-recovery criterion rather than a pointwise regression loss. 
Fig.\ref{fig:graphunwrapnet} provides a visual representation of the primary computational flow and internal modules that constitute our proposed framework.

\section{Experimental Results}

\label{sec:exp}
\subsection{Datasets and Preprocessing}
We evaluate the proposed method on two 14-channel EEG datasets acquired with Emotiv devices. The first is STEW \cite{lim2018STEW}, which contains workload-related EEG recordings from 48 subjects acquired using the Emotiv EPOC headset. Following prior modulo EEG recovery work \cite{geng2023transformer}, the raw recordings are partitioned into non-overlapping segments of 200 samples. The second is MIMED \cite{wirawan2024mimed}, recorded with the Emotiv EPOC X from 30 participants, and containing motor execution and motor imagery sessions for right-hand up/down, left-hand up/down, standing, and sitting.

Each subject is normalized independently using the median/MAD sigmoid mapping in ~\eqref{eq:robust_norm} for both datasets. Synthetic modulo folding creates wrapped observations. The wrapped signal is converted to the phase-aware feature representation in \eqref{eq:feat}. To test robustness under increased dynamic-range compression, experiments are conducted at mild to severe folding thresholds.

\subsection{Training and Evaluation Protocol}

For each dataset, we use subject-wise train/validation/test partitions and keep the same split across all folding thresholds for fair comparison. The model is trained using AdamW with learning rate $2\times 10^{-4}$, weight decay $5\times 10^{-4}$, batch size $64$, and $100$ epochs. A cosine annealing schedule is used for learning rate decay, and gradient clipping is applied to stabilize optimization. Training is implemented in PyTorch. The architectural and optimization hyperparameters are summarized in Table~\ref{tab:implementation}.
Performance is evaluated using fold-state accuracy, mean absolute error, mean squared error, and Pearson correlation coefficient. 
Fold-state accuracy measures discrete recovery quality, while the signal-domain metrics assess waveform fidelity and temporal agreement with the ground truth. 
We further study the proposed framework through component ablations and additive-noise experiments.

\begin{table}[t]
\centering
\caption{Implementation hyperparameters.}
\label{tab:implementation}
\begin{tabular}{lc}
\hline
\textbf{Hyperparameter} & \textbf{Value} \\
\hline
Segment length, $T$                 & 200 \\
Graph neighbors, $\kappa$          & 3 \\
Maximum fold jump, $K$             & 4 \\
Hidden dimension, $H$              & 96 \\
Graph-temporal layers, $L$         & 6 \\
Dropout rate                        & 0.10 \\
Potts penalty, $\beta$              & 0.03 \\
Residual scale, $\rho$              & 0.03 \\
FiLM hidden dimension              & 64 \\
FiLM scale, $\alpha_f$              & 0.05 \\
CRF loss weight, $w_{\mathrm{crf}}$ & 1.00 \\
Increment loss weight, $w_k$       & 0.25 \\
Signal loss weight, $w_x$          & 0.25 \\
Difference loss weight, $w_{\Delta x}$ & 0.15 \\
\hline
\end{tabular}
\end{table}

\subsection{Benchmark Comparison}

\begin{figure*}
    \centering
    \subfigure[Reconstruction for $\lambda = 0.6$.]{\includegraphics[width=0.7\textwidth]{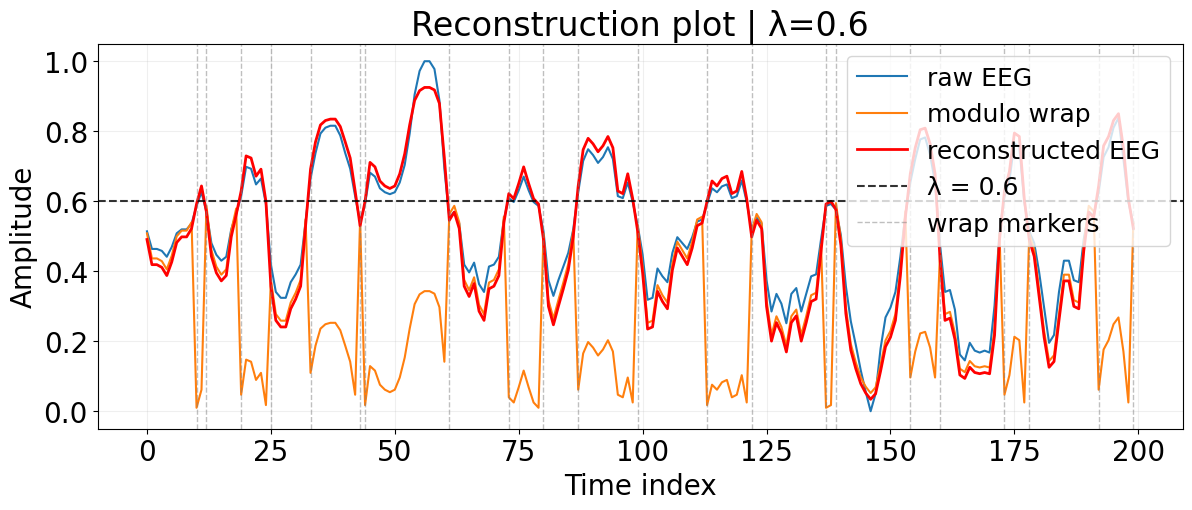}}  \\
    \subfigure[Reconstruction for $\lambda = 0.5$.]{\includegraphics[width=0.7\textwidth]{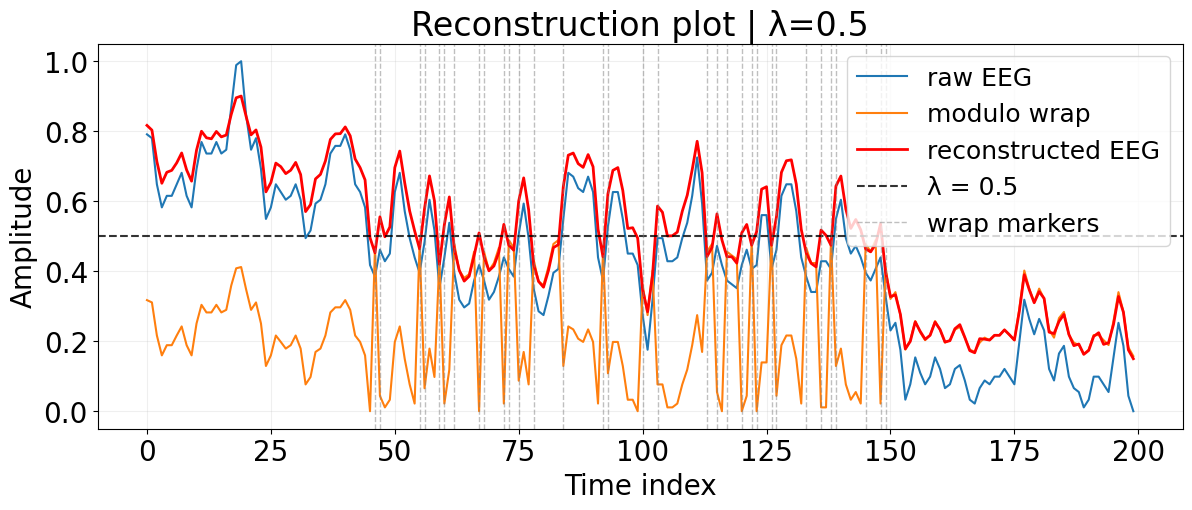}} \\
    \subfigure[Reconstruction for $\lambda = 0.4$.]{\includegraphics[width=0.7\textwidth]{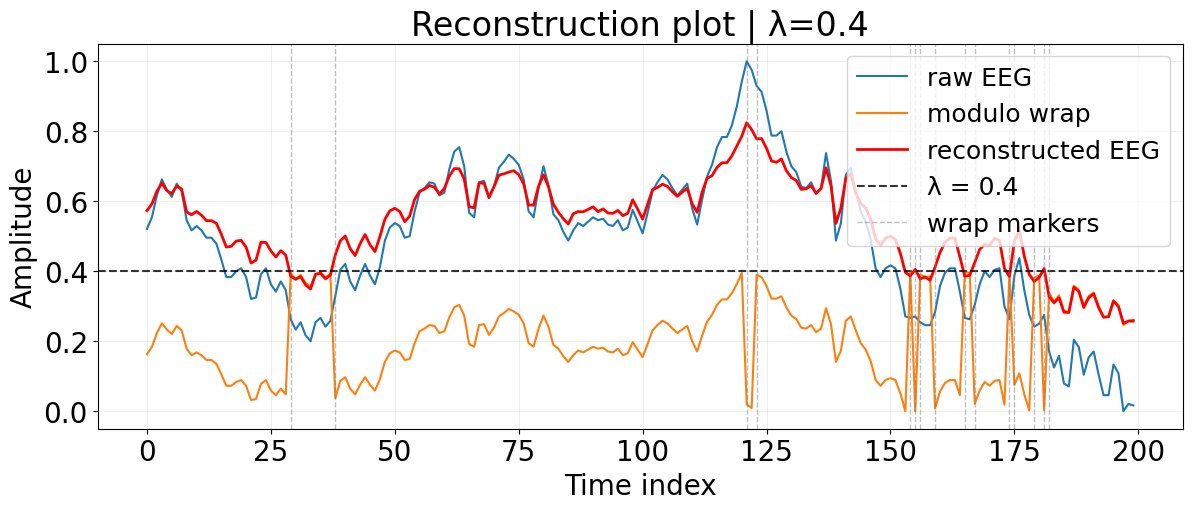}}
    \caption{Reconstruction examples of the proposed method under different folding thresholds. Blue: raw EEG; orange: modulo observation; red: reconstructed EEG. The plots illustrate that the proposed method preserves the main temporal structure.}
    \label{fig:recon_examples}
\end{figure*}

Table~\ref{tab:stew_comp} reports our evaluation on STEW~\cite{lim2018STEW} at different folding thresholds. Our proposed method achieves the highest fold-state accuracy and correlation across all three thresholds, indicating more accurate retrieval of the latent fold trajectory. The accuracy gain increases with folding strength, consistent with organized decoding under increased ambiguity. Waveform errors are not repeatedly lowest, but enhanced fold-state inference leads to better temporal agreement with the ground truth. These results support modeling modulo EEG recovery as structured fold-state estimation rather than pointwise regression.
Fig.~\ref{fig:recon_examples} shows reconstruction examples of the proposed method on STEW~\cite{lim2018STEW} under three folding thresholds. 
As the folding threshold decreases, the reconstruction becomes more challenging because wrap ambiguity increases, but the proposed method still recovers the main waveform trend with good fidelity.



\begin{table*}[t]
\caption{Comparison results on STEW~\cite{lim2018STEW} with MRF~\cite{zhao2015MRF}, Sparse Opt.~\cite{feng2022SparseOptimization}, Transformer~\cite{geng2023transformer}, and B2R2~\cite{azar2022residual} methods for different folding thresholds ($\lambda$).}
\label{tab:stew_comp}
\centering
\renewcommand{\arraystretch}{0.95}
\begin{tabular}{c l c c c c}
\toprule
$\lambda$ & Method & Acc$_z$ & L1 & MSE & $R$ \\
\midrule
0.6 & MRF         & 75.85 & 0.14 & 0.08 & 0.17 \\
    & Sparse Opt. & 76.68 & 0.14 & 0.08 & 0.16 \\
    & Transformer & 85.86 & 0.09 & 0.04 & 0.34 \\
    & B2R2        & 87.98 & \textbf{0.07} & \textbf{0.04} & 0.62 \\
    & Proposed    & \textbf{92.01} & 0.08 & 0.05 & \textbf{0.84} \\
\midrule
0.5 & MRF         & 50.77 & 0.24 & 0.12 & 0.13 \\
    & Sparse Opt. & 61.77 & 0.25 & 0.12 & 0.11 \\
    & Transformer & 78.96 & 0.13 & \textbf{0.04} & 0.16 \\
    & B2R2        & 79.89 & \textbf{0.11} & \textbf{0.04} & 0.44 \\
    & Proposed    & \textbf{83.31} & \textbf{0.11} & 0.05 & \textbf{0.76} \\
\midrule
0.4 & MRF         & 32.53 & 0.28 & 0.12 & 0.07 \\
    & Sparse Opt. & 44.38 & 0.28 & 0.11 & 0.12 \\
    & Transformer & 57.68 & 0.16 & \textbf{0.05} & 0.01 \\
    & B2R2        & 59.12 & \textbf{0.14} & 0.09 & 0.34 \\
    & Proposed    & \textbf{71.56} & \textbf{0.14} & 0.06 & \textbf{0.68} \\
\bottomrule
\end{tabular}
\end{table*}

\begin{table*}
\caption{Proposed method across folding thresholds on STEW~\cite{lim2018STEW} and MIMED~\cite{wirawan2024mimed}.}
\label{tab:robustness}
\centering
\begin{tabular}{c|cccc|cccc}
\hline
& \multicolumn{4}{c|}{STEW~\cite{lim2018STEW}} & \multicolumn{4}{c}{MIMED~\cite{wirawan2024mimed}} \\
\hline
$\lambda$ & Acc$_z$ & L1 & MSE & $R$ & Acc$_z$ & L1 & MSE & $R$ \\
\hline
0.6  & 92.01 & 0.08 & 0.05 & 0.84 & 89.35 & 0.21 & 0.06 & 0.92 \\
0.5  & 83.31 & 0.11 & 0.05 & 0.76 & 83.24 & 0.25 & 0.10 & 0.80 \\
0.4  & 71.56 & 0.14 & 0.06 & 0.68 & 81.29 & 0.25 & 0.08 & 0.79 \\
0.3  & 60.04 & 0.18 & 0.10 & 0.60 & 70.12 & 0.26 & 0.09 & 0.74 \\
0.25 & 54.94 & 0.28 & 0.11 & 0.65 & 61.45 & 0.26 & 0.09 & 0.71 \\
0.2  & 44.77 & 0.23 & 0.09 & 0.58 & 50.94 & 0.26 & 0.09 & 0.69 \\
\hline
\end{tabular}
\end{table*}


\begin{table*}
\caption{Ablation study on STEW~\cite{lim2018STEW}. Res = residual refinement; GMix = graph mixing. 
}
\label{tab:ablation}
\centering
\begin{tabular}{l|cc|cc|cc}
\hline
& \multicolumn{2}{c|}{$\lambda=0.6$} & \multicolumn{2}{c|}{$\lambda=0.5$} & \multicolumn{2}{c}{$\lambda=0.4$} \\
\hline
Variant & Acc$_z$ & $R$ & Acc$_z$ & $R$ & Acc$_z$ & $R$ \\
\hline
Full                       & \textbf{92.01} & \textbf{0.84} & \textbf{83.31} & \textbf{0.76} & \textbf{71.56} & \textbf{0.68} \\
w/o Potts                  & 91.34          & 0.81          & 81.77          & 0.75          & 70.23          & 0.65 \\
w/o Potts + PGFI           & 89.68          & 0.78          & 78.85          & 0.72          & 68.87          & 0.61 \\
w/o CRF                    & 66.95          & 0.68          & 57.47          & 0.55          & 50.52          & 0.43 \\
w/o CRF + Res              & 65.03          & 0.48          & 55.90          & 0.45          & 49.12          & 0.32 \\
w/o CRF + Res + FiLM       & 63.21          & 0.43          & 53.54          & 0.43          & 47.06          & 0.30 \\
w/o CRF + Res + FiLM + GMix& 60.17          & 0.42          & 50.23          & 0.40          & 45.89          & 0.27 \\
\hline
\end{tabular}%
\end{table*}

\subsection{Robustness Across Thresholds and Datasets}

Table~\ref{tab:robustness} reports the proposed method across a wider range of folding thresholds on STEW~\cite{lim2018STEW} and MIMED~\cite{wirawan2024mimed}. All results are obtained under our evaluation pipeline. As the folding threshold decreases, performance drops on both datasets, which is expected due to increased ambiguity in the hidden fold-state sequence. Even under severe folding, the method maintains meaningful correlation with the ground truth. MIMED~\cite{wirawan2024mimed} remains more stable across thresholds. STEW~\cite{lim2018STEW} shows stronger degradation at lower thresholds.

Table~\ref{tab:noise_full} reports the noise-robustness study on STEW~\cite{lim2018STEW} under corruptions. In addition to additive and colored noise, we consider structured interference, physiological artifact, missing data, and composite perturbations. Here, \emph{channel dropout} denotes random masking of whole EEG channels, \emph{temporal dropout} denotes random masking of time samples, and \emph{composite} denotes a chained corruption formed by sequentially applying multiple noise types.
Performance is relatively stable under mild perturbations, while larger degradation appears under structured and dropout-type corruptions. 


\begin{table*}[t]
\centering
\caption{Noise-robustness results on STEW~\cite{lim2018STEW} under the tested corruptions.}
\label{tab:noise_full}

\begin{tabular}{c l l c c}
\hline
$\lambda$ & Type & Parameters & Acc$_z$ & $R$ \\
\hline

0.4 & none & --- & 71.56 & 0.680 \\
0.4 & gauss & $\sigma=0.01$ & 70.45 & 0.678 \\
0.4 & gauss & $\sigma=0.05$ & 54.88 & 0.535 \\
0.4 & pink & $\sigma=0.03$ & 68.48 & 0.674 \\
0.4 & brown & $\sigma=0.03$ & 69.11 & 0.680 \\
0.4 & line 60 Hz & amp $=0.03$ & 59.31 & 0.560 \\
0.4 & drift & amp $=0.06$, $f_{\max}=0.5$ & 69.46 & 0.677 \\
0.4 & emg & $\sigma=0.03$, $f_{\mathrm{lo}}=20$, $f_{\mathrm{hi}}=45$ & 65.16 & 0.639 \\
0.4 & impulse & prob $=0.002$, amp $=0.5$ & 71.40 & 0.679 \\
0.4 & channel drop & $p=0.15$, fill $=0.5$ & 65.26 & 0.589 \\
0.4 & temporal drop & $p=0.05$, fill $=0.5$ & 64.08 & 0.615 \\
0.4 & composite &
gauss($\sigma=0.02$) $+$ line 50 Hz(amp $=0.02$) $+$ drift(amp $=0.04$, $f_{\max}=0.4$)
& 64.42 & 0.627 \\

\hline

0.5 & none & --- & 83.31 & 0.760 \\
0.5 & gauss & $\sigma=0.01$ & 78.26 & 0.705 \\
0.5 & gauss & $\sigma=0.05$ & 63.50 & 0.528 \\
0.5 & pink & $\sigma=0.03$ & 74.65 & 0.682 \\
0.5 & brown & $\sigma=0.03$ & 77.68 & 0.746 \\
0.5 & line 60 Hz & amp $=0.03$ & 71.41 & 0.623 \\
0.5 & drift & amp $=0.06$, $f_{\max}=0.5$ & 78.74 & 0.744 \\
0.5 & emg & $\sigma=0.03$, $f_{\mathrm{lo}}=20$, $f_{\mathrm{hi}}=45$ & 71.31 & 0.627 \\
0.5 & impulse & prob $=0.002$, amp $=0.5$ & 83.07 & 0.758 \\
0.5 & channel drop & $p=0.15$, fill $=0.5$ & 76.81 & 0.603 \\
0.5 & temporal drop & $p=0.05$, fill $=0.5$ & 70.16 & 0.604 \\
0.5 & composite &
gauss($\sigma=0.02$) $+$ line 50 Hz(amp $=0.02$) $+$ drift(amp $=0.04$, $f_{\max}=0.4$)
& 75.90 & 0.688 \\

\hline

0.6 & none & --- & 92.01 & 0.840 \\
0.6 & gauss & $\sigma=0.01$ & 91.41 & 0.839 \\
0.6 & gauss & $\sigma=0.05$ & 84.45 & 0.755 \\
0.6 & pink & $\sigma=0.03$ & 90.45 & 0.838 \\
0.6 & brown & $\sigma=0.03$ & 90.41 & 0.838 \\
0.6 & line 60 Hz & amp $=0.03$ & 87.04 & 0.781 \\
0.6 & drift & amp $=0.06$, $f_{\max}=0.5$ & 90.68 & 0.837 \\
0.6 & emg & $\sigma=0.03$, $f_{\mathrm{lo}}=20$, $f_{\mathrm{hi}}=45$ & 89.73 & 0.827 \\
0.6 & impulse & prob $=0.002$, amp $=0.5$ & 91.39 & 0.830 \\
0.6 & channel drop & $p=0.15$, fill $=0.5$ & 86.11 & 0.768 \\
0.6 & temporal drop & $p=0.05$, fill $=0.5$ & 75.65 & 0.581 \\
0.6 & composite &
gauss($\sigma=0.02$) $+$ line 50 Hz(amp $=0.02$) $+$ drift(amp $=0.04$, $f_{\max}=0.4$)
& 88.84 & 0.812 \\

\hline
\end{tabular}
\end{table*}


\subsection{Ablation Study}
Table~\ref{tab:ablation} reports an ablation study on STEW~\cite{lim2018STEW}. 
The ablation variants remove the following components from the full model: \emph{i)} the Potts prior in the CRF transition score, \emph{ii)} the PGFI gate used for temperature-scaled transition sharpening, \emph{iii)} CRF-based structured fold-state decoding, \emph{iv)} the residual refinement term in the reconstruction equation, \emph{v)} FiLM-style hidden-feature calibration, and \emph{vi)} GraphMix across channels.
The Potts prior is the penalty term in ~\eqref{eq:transition}, where nonzero fold-state transitions incur an additional cost controlled by $\beta$. The PGFI gate is the boundary-guided scalar $g_{t,c}$ in ~\eqref{eq:temp}, which sharpens transition evidence near likely wrap events. GraphMix refers to the graph propagation step in ~\eqref{eq:backbone}, where the normalized adjacency operator mixes hidden features across neighboring electrodes at each time index. Residual refinement corresponds to the bounded correction term $r_{t,c}$ in ~\eqref{eq:recon_hard} and ~\eqref{eq:recon_soft}, which compensates for small reconstruction bias after fold-state estimation.
For FiLM calibration, let $\boldsymbol{\mu}\in\mathbb{R}^{C}$ and $\boldsymbol{\sigma}\in\mathbb{R}^{C}$ denote the channel-wise mean and standard deviation of the wrapped segment $\mathbf{p}$. We form
$\mathbf{s}=[\boldsymbol{\mu};\boldsymbol{\sigma}],$
and use a small multilayer perceptron to predict affine parameters $(\boldsymbol{\gamma},\boldsymbol{\delta})$. The hidden representation is then modulated as
\begin{equation}
\mathbf{h}\leftarrow (1+\alpha_f\boldsymbol{\gamma})\odot \mathbf{h}+\alpha_f\boldsymbol{\delta},
\end{equation}
where $\alpha_f$ is a small scaling factor and $\odot$ denotes elementwise multiplication. This calibration improves robustness to variation across subjects and recording conditions.

The full model performs best at all three thresholds. Removing the Potts prior and boundary gate both reduce performance, demonstrating that transition regularization and boundary-aware sharpening help recover. Structured fold-state decoding dominates when the CRF is removed, resulting in deterioration. Residual refinement, FiLM calibration, and graph mixing decrease correlation after removal, indicating their complementary roles in waveform recovery and cross-channel modeling.

\subsection{Computational Complexity}
\label{subsec:complexity}

Let $B$, $C$, $T$, $H$, $L$, $q$, and $S$ denote the batch
size, number of channels, segment length, hidden dimension,
number of graph-temporal layers, temporal kernel width, and
number of admissible fold states, respectively. The dominant
backbone complexity is
\[
\mathcal{O}\!\left(BLCTqH^2+BLC^2TH\right),
\]
where the first term corresponds to the channel-wise temporal
convolutions and the second to the dense GraphMix operation.
The remaining prediction heads have lower-order complexity.
The implemented CRF constructs a masked dense $S\times S$
transition tensor. Consequently, Viterbi decoding requires
$\mathcal{O}(BCTS^2)$ operations and $\mathcal{O}(BCTS)$
backpointer storage. Although transitions satisfying
$|j-i|>K$ are excluded from valid paths, they are masked after
the dense transition tensor is formed.
With $C=14$, $T=200$, $H=96$, $L=6$, and $q=3$, the complete
model contains $188{,}882$ trainable parameters, corresponding
to approximately $0.72$ MiB of FP32 parameter storage. The
convolutional and linear operations require approximately
$506.0$ MMACs per segment, of which the temporal convolutions
account for approximately $92\%$. For the largest state space
used by the current implementation ($S=6$), Viterbi decoding
evaluates approximately $1.0\times10^5$ transition scores per
segment.


\section{Conclusion}
\label{sec:conc}

We presented a signal-structured framework for modulo EEG recovery that formulates reconstruction as estimation of a discrete fold-state trajectory. The proposed method combines phase-aware wrapped-domain features, graph-temporal representation learning, and CRF-based decoding to exploit modulo periodicity, electrode dependence, and temporal consistency. Boundary-aware transition sharpening, FiLM calibration, and bounded residual refinement further improve reconstruction near ambiguous wrap events and across varying recording conditions.

Experiments on STEW~\cite{lim2018STEW} and MIMED~\cite{wirawan2024mimed} demonstrated consistent recovery across mild-to-severe folding thresholds. In particular, the improvements in fold-state accuracy and waveform correlation indicate that globally structured decoding can suppress isolated transition errors that otherwise produce large unfolding offsets. The ablation study confirmed the dominant contribution of the CRF decoder, while the robustness experiments showed that the framework remains stable under several common noise and missing-data perturbations.

Future work will consider uncertain or time-varying folding thresholds, acquisition non-idealities, and threshold-independent recovery models. Real-time implementation under streaming constraints, validation on physical modulo acquisition hardware, and evaluation through downstream EEG tasks will also be important for establishing practical utility. Extending the structured recovery formulation to other multichannel biomedical signals is another promising direction.




\printcredits

\bibliographystyle{cas-model2-names}

\bibliography{cas-refs}





\end{document}